\documentclass[usenatbib]{mnras}
\usepackage[utf8]{inputenc}
\usepackage{graphicx}
\usepackage{caption}
\usepackage{subcaption}
\usepackage{amsmath}
\usepackage{authblk}
\usepackage{cuted}

\title[Atmospheric Shocks]{The Propagation of Strong Shocks into Planetary and Stellar Atmospheres}

\author[A. Yalinewich and A. Remorov]{
Almog Yalinewich,$^{1}$
Andrey Remorov,$^{1}$

$^{1}$Canadian Institute for Theoretical Astrophysics, 60 St. George St., Toronto, ON M5S 3H8, Canada
}

\date{Accepted XXX. Received YYY; in original form ZZZ}

\pubyear{2021}

\begin{document}
\label{firstpage}
\pagerange{\pageref{firstpage}--\pageref{lastpage}}
\maketitle

\begin{abstract}
In this work we present a mathematical model for the propagation of the shock waves that occur in graded density profiles. These waves can occur in a wide range of astrophysical events, such as collisions in planetary and stellar atmospheres, common envelope explosions and peculiar type Ia supernovae. The behaviour of the shock wave and its evolution can be modelled using type II self similar solutions. In such solutions the evolution of the shock wave is determined by boundary conditions at the shock front and a singular point in the shocked region. We show how the evolution can be determined for different equations of state and density profiles, and compare these results to numerical simulations. These findings are also applied to a variety of astrophysical phenomena to further test their validity.
\end{abstract}




\section{Introduction}

The problem of a surface explosion has been considered by many authors in the past, using numerical simulations \citep{Potter2012ConstrainingImpact, Monteux2019ShockRevisited} and lab experiments \citep{Holsapple2012MomentumScaling, Mazzariol2019ExperimentalMaterials, Holsapple1987ThePhenomena}.
The primarily incentive for these works was to study cratering events due to planetary impacts. Among the terrestrial impact events considered are the Chicxulub Cretaceous/Tertiary impact \citep{Pope1997EnergyImpact, Newman1999ImpactResults} and the Tunguska event \citep{Artemieva2016FromJupiter}. Related phenomena occurring on other astronomical bodies are being explored as well. Some of these include asteroid impacts on the atmosphere of Venus \citep{Korycansky2000High-ResolutionAtmosphere}, meteoroid collisions causing particle ejections from the Bennu asteroid \citep{Bottke2020MeteoroidEvents}, and various impacting objects colliding with Jupiter causing flashes of certain light curves \citep{Hueso2013ImpactCollisions}, the most famous of which is the light curve generated by the Shoemaker Levy 9 collision \citep{Zahnle1995AImpact}. That being said, subsurface explosions are not restricted to planets, and can also occur in star. For example, it has been suggested that the 1954 precursor to the peculiar supernova iptf2014hls was caused by an explosion in a common envelope \citep{Yalinewich2019OpticalSurface}.

In this work, we develop an analytic formalism to model the propagation of strong shocks into an increasing density profiles, such as planetary and stellar atmospheres. For simplicity, we will assume that the explosion is so strong that we can neglect the ambient pressure, and that the radius is much larger than the size of the initial hot spot. We will also neglect gravity and assume an ideal gas equation of state for all materials. Under these assumptions, we can use the method of self similarity to describe the evolution of the shock wave \citep{Barenblatt2014ScalingAsymptotics}.  The propagation of such a shock wave will be markedly different from the propagation of a shock wave into a medium with flat or shallow density profile. The latter can be described by the celebrated Sedov Taylor solution \citep{Sedov1946PropagationWaves, Taylor1950TheDiscussion}, where behaviour of the shock is governed by energy conservation. In contrast, in the problem we discuss the energy available to drive the shock decreases with time. This is because shocked fluid elements expand and accelerate away from the shock front, and at some point cross a critical position called the sonic point where information cannot travel back to the shock front. For this reason, the behaviour of this shock is determined by the behaviour near this sonic point. Such solutions to the hydrodynamic equations are termed ``type II'', to distinguish them from the ``type I'' solutions, which are governed by energy conservation. These type II solutions have previously been used to model accelerating shocks propagating into a steeply declining density profiles, \citep{Waxman1992Second-typeProblem, Schlichting2014FormationImplications}. In contrast, in this work we consider a decelerating type II shocks propagating into an increasing density profile.

Our model is a novel generalisation to the classical impulsive piston problem \citep{Adamskii1956IntegrationGas, Yalinewich2020SelfCollisions}. In the classical problem, a thin wafer collides with a semi infinite space filled with a cold gas with a uniform density. One can consider this problem as the one dimensional analogue of an impact event. Since this problem is one dimensional and slab symmetric, it can be solved using the self similar method. The solution predicts a power law relation between the shock velocity and the swept up mass $v \propto m^{-\beta}$. Interestingly, the same power law relation also holds well in the three dimensional case (i.e. a sphere hitting a flat surface head on). In this study we consider a variation on the classical problem where the density of target can vary as some power law of distance from the impact site $\rho \propto x^{\omega}$. Not only will this description allow us to consider an atmospheric graded density profile, but it will also help us understand why the one dimensional analogue describes the three dimensional case well. This is because one way to think about the difference between the slab symmetric and three dimensional cases is how the mass increases with the distance. For example, if the density is uniform, then in slab geometry the mass increases linearly with the distance while in three dimensions the mass increases as a cube of the distance. These differences can be accounted for in the slab symmetric analogue by changing the density profile in the target. 

The plan of the paper is as follows. In section \ref{sec:ipp} we show how self similarity can be used to solve the generalised impulsive piston problem. In section \ref{sec:app} we demonstrate how the solutions we found can be used to model a variety of astrophysical problems. In section \ref{sec:conclusion} we discuss the results and conclude.

\section{Self Similar Model} \label{sec:ipp}

\subsection{Impulsive Piston Problem}

Let us consider an explosion close to the surface of a large star or planet, so that the radius of curvature is unimportant and the atmospheric layers can be considered planar. This explosion will generate a shock wave that travels into the atmosphere. While the distance between the centre of the shock and the shock front is much smaller than the atmospheric scale height, the shock front interacts with an ambient medium with the same density, and hence the behaviour can be described by the classical Sedov Taylor explosion \citep{Sedov1946PropagationWaves, Taylor1950TheDiscussion}.
When the distance between the shock front and explosion centre becomes comparable to the scale height, the top of the shock encounters a declining density profile and accelerates. A mathematical model describing this part of the shock was developed in \citep{Sakurai1960OnGas}. In this work we focus on the part of the shock that propagates deeper into the atmosphere, and since the density the shock encounters increases the shock decelerates. When the distance between the centre of the explosion and the deepest point on the shock front is much larger than the atmospheric scale height the shock becomes self similar. Therefore, the effective radius of the shock grows as some power law of time $R \propto t^{\alpha}$, where
We define the effective radius of the shock as the distance between the deepest point of the shock front to the centre of the explosion.

As was done in a previous paper \citep{Yalinewich2020SelfCollisions}, instead of considering a three dimensional problem, we consider the one-dimensional analogue called the impulsive piston problem \citep{ Adamskii1956IntegrationGas, Zeldovich1967PhysicsPhenomena}. In this scenario, a thin wafer hits a much thicker slab of material (with a certain density distribution) and both are perfectly cold prior to the collision. As a result of this collision a shock wave emerges from the contact surface and travels into the target. Once the shock wave has travelled a distance much larger than the width of the wafer, there is no other relevant length scale in this process other than the position of the shock front, and hence we expect the shock wave to evolve in a self similar way. In the original one dimensional impulsive piston problem, the target is considered to have uniform density. However, in this work we consider a graded density profile defined by a power law dependence on the distance from the edge
\begin{equation}
    \rho_a = k x^{\omega}
\end{equation}
where $k$ and $\omega>0$ are constants and $x$ is the distance from the edge.

In the next sections we describe how self similarity allows us to solve the impulsive piston problem. It does so by reduce the hydrodynamic equations, which are partial differential equations, to ordinary differential equations. We also describe how these equation can be numerically integrated, and how the value of $\alpha$ can be obtained, as was done for the case $\omega=0$ in \citep{Yalinewich2020SelfCollisions}.

\subsection{Self Similar Equations}

The mathematical machinery presented in the following subsections is very similar to the methods used to analyse cratering events in planetary collisions in a previous paper \citep{Yalinewich2020SelfCollisions}. The main difference is that the previous work considered a uniform density profile, compared to the graded density profile examined here.

The slab symmetric hydrodynamic equations in one dimension are given by \citep{Landau1987FluidMechanics}, which are respectively the continuity, momentum, and entropy conservation, where $\rho$ is density, $v$ is velocity, $p$ is pressure, and $s$ is entropy:

\begin{equation}
\frac{\partial \rho }{\partial t}+ 
\rho \frac{\partial v }{\partial x} + v \frac{\partial \rho}{\partial x} = 0 \label{eq:mass_conservation}
\end{equation}

\begin{equation}
\frac{\partial v}{\partial t} +  v\frac{\partial v}{\partial x}+ 
\frac{1}{\rho} \frac{\partial p}{\partial x}  = 0 \label{eq:momentum_conservation}
\end{equation}

\begin{equation}
\frac{\partial s}{\partial t} + v \frac{\partial s}{\partial x}= 0 \, . \label{eq:entropy_conservation}
\end{equation}
We assume an ideal gas equation of state
\begin{equation}
s = \ln p - \gamma \ln \rho
\end{equation}
where $\gamma$ is the adiabatic index.

We can replace the pressure by the sound speed using the relation $p =  \rho c^2 / \gamma$. After this substitution we have a system of partial differential equations in $v, c$ and $\rho$. We make the assumption that the one dimensional shock has a self similar behaviour, so the position of the shock front $X(t)$ evolves as a power law in time
\begin{equation}
    X(t) \propto t^\alpha \, . \label{eq:shock_power_law}
\end{equation}
where $\alpha$ is a constant. One of the main goals of the analysis is determining $\alpha$. The partial differential equations can be reduced to ordinary differential equations in the dimensionless position  $\chi = x/X$. Following \citep{Landau1987FluidMechanics}, we define dimensionless hydrodynamical variables $V, C, D$ in the following way:
\begin{equation}
    v(x,t) = \frac{d X(t)}{dt}\chi  V\left(\chi\right) 
\end{equation} 
\begin{equation}
    c(x,t) = \frac{d X(t)}{dt} \chi C\left(\chi\right)
\end{equation} 
\begin{equation}
    \rho(x,t) = k X^{\omega}(t) D\left(\chi\right) \, . \label{eq:self_similar_vars}
\end{equation}
After this substitution, the equations contain terms involving different time derivatives of $X$. These time derivatives can be factored out using equation \ref{eq:shock_power_law}. We introduce the following parameter
\begin{equation}
    \delta = \frac{\ddot{X} X}{\dot{X}^2} = 1-\frac{1}{\alpha} \,.\label{eq:X_derivatives}
\end{equation}
This parameter is the power law index of the shock velocity - position relation $d X/dt \propto X^{\delta}$.

After making these adjustments and simplifying, the hydrodynamic equations equations \ref{eq:mass_conservation}, \ref{eq:momentum_conservation} and \ref{eq:entropy_conservation}
become the matrix equation $\overleftrightarrow{M} d\vec{A}/\chi = \vec{B}$ where:

\begin{equation}
M = \begin{bmatrix}\chi \left(V - 1\right) & \chi D & 0\\\chi C^{2} & \gamma \chi (V - 1) D & 2 \chi C D\\ C \chi  (\gamma - 1) \left(1-V\right) & 0 & 2 D \chi \left(V - 1\right) \end{bmatrix}
\end{equation}

\begin{equation}
\vec{A} = \begin{bmatrix} D\\ V\\ C\end{bmatrix}
\end{equation}

\begin{equation}
    \vec{B}= \begin{bmatrix}-\omega D - D V \\- \delta \gamma D V - \gamma (V^{2} - V) D - 2 C^{2} D\\ -(-\gamma \omega + \omega) C D - 2 C D V\end{bmatrix}
\end{equation}

These equations are linear in the derivatives $dV/d\chi, dC/d\chi, dD/d\chi$ so these can be isolated. Our choice of parameters allows us to go a step further and obtain a single ordinary differential equation involving just $C$ and $V$. We note that

\begin{equation}
\frac{d V}{d\chi} = \frac{\tilde{N}_1}{\tilde{D}_1} \label{eq:dVdx}
\end{equation}

and

\begin{equation}
\frac{d C}{d \chi} = \frac{N_1}{D_1}
\end{equation}
where
\begin{equation}
    \tilde{N}_1 = - C^{2} V \gamma - 2 C^{2} \delta + V^{3} \gamma + V^{2} \delta \gamma - 2 V^{2} \gamma - V \delta \gamma + V \gamma - \omega C^2
\end{equation}
\begin{equation}
    \tilde{D}_1 = \chi \gamma \left(C^{2} - V^{2} + 2 V - 1\right)
\end{equation}
\begin{strip}
\begin{equation*}
N_1 = C \left((V + \omega) \gamma \left(V - 1\right) \left(V \gamma - V - \gamma + 1\right) - \left(C^{2} - \gamma \left(V - 1\right)^{2}\right)\times \right.
\end{equation*}
\begin{equation}
 \left. \times\left(2V + 2\delta - \omega \gamma + \omega - 2\right) - \left(2 C^{2} + V^{2} \gamma + V \delta \gamma - V \gamma\right) \left(V \gamma - V - \gamma + 1\right)\right)
\end{equation}
\begin{equation}
D_1 = 2 \chi \left(C^{2} \left(V - 1\right) + C^{2} \left(V \gamma - V - \gamma + 1\right) - \gamma \left(V - 1\right)^{3}\right)
\end{equation}
\end{strip}
Dividing one equation by the other, we get a single ODE

\begin{equation}
\frac{dC}{d V} = \frac{N_2}{D_2} \label{eq:dCdV_raw}
\end{equation}
where:
\begin{equation*}
    N_2 = C \gamma \left(- (V + \omega) \gamma \left(V - 1\right) \left(V \gamma - V - \gamma + 1\right) +\right.
\end{equation*}
\begin{equation*}
    \left. \left(C^{2} - \gamma \left(V - 1\right)^{2}\right) \left(2V + 2\delta -\gamma \omega + \omega - 2\right) + \right.
\end{equation*}
\begin{equation}
    \left. + \left(2 C^{2} + V^{2} \gamma + V \delta \gamma - V \gamma\right) \left(V \gamma - V - \gamma + 1\right)\right) \times  \label{eq:dCdV_num}
\end{equation}
\begin{equation*}
    \left(C^{2} - V^{2} + 2 V - 1\right)
\end{equation*}

and
\begin{equation}
D_2 = 2 \left(C^{2} \left(V - 1\right) + C^{2} \left(V \gamma - V - \gamma + 1\right) - \gamma \left(V - 1\right)^{3}\right) \times  \label{eq:dCdV_den}
\end{equation}
\begin{equation*}
    \times \left(C^{2} V \gamma + 2 C^{2} \delta + C^2 \omega- V^{3} \gamma - V^{2} \delta \gamma + 2 V^{2} \gamma + V \delta \gamma - V \gamma\right) \, .
\end{equation*}

\subsection{Boundary Conditions}

In order to integrate equation \ref{eq:dCdV_raw}, boundary conditions and the value of $\delta$ are needed. The boundary conditions at the shock front are given by the Rankine Hugoniot conditions for a strong shock \citep{Zeldovich1967PhysicsPhenomena, Landau1987FluidMechanics}.
\begin{equation}
    V_f = \frac{2}{\gamma+1} \label{eq:v_shock_front}
\end{equation}
and
\begin{equation}
    C_f = \frac{\sqrt{2 \gamma \left(\gamma-1\right)}}{\gamma+1} \, .  \label{eq:C_final}
\end{equation}
In some self similar problems, like the Sedov Taylor explosion, the parameter $\delta$ can be be inferred directly from conservation laws. Such problems are known as Type I solutions. In our case, however, $\delta$ is determined by the condition that the hydrodynamic trajectory passes smoothly through some singularity. These are called Type II solutions \citep{Waxman1992Second-typeProblem}. The singularity occurs at a point where information cannot propagate back to the shock front, and is therefore also referred to as the sonic point. Energy that flows through the sonic point cannot travel back to the shock front, and so the energy available to sustain the shock is not conserved.

Equation \ref{eq:dVdx} has a singularity when 
\begin{equation}
C = 1- V \label{eq:sonic_line}
\end{equation}
This curve is known as the sonic line. On the sonic line, the denominator (equation \ref{eq:dCdV_den}) vanishes. To prevent a divergence, the numerator also has to vanish. This happens on a specific point on the sonic line, which we call the sonic point

\begin{equation}
V_s = \frac{2\delta + \omega}{\delta (2-\gamma)+\omega} \label{eq:v_sonic}
\end{equation}
\begin{equation}
C_s = - \frac{\delta \gamma}{\delta (2- \gamma) + \omega} \label{eq:c_sonic}
\end{equation}

Knowing the boundary conditions and $\frac{d C}{d V}$, we can integrate from the sonic point to the shock front. To avoid a numerical divergence, the starting point is slightly shifted from the sonic point in the positive $V$ direction by $dV_i$, and in $C$ direction by $d V_i \frac{d C}{d V}\Bigr|_{sonic}$ for some small $d V_i \ll 1$. 

\begin{figure}
\centering
\begin{subfigure}{.5\textwidth}
  \centering
  \includegraphics[width=1\linewidth]{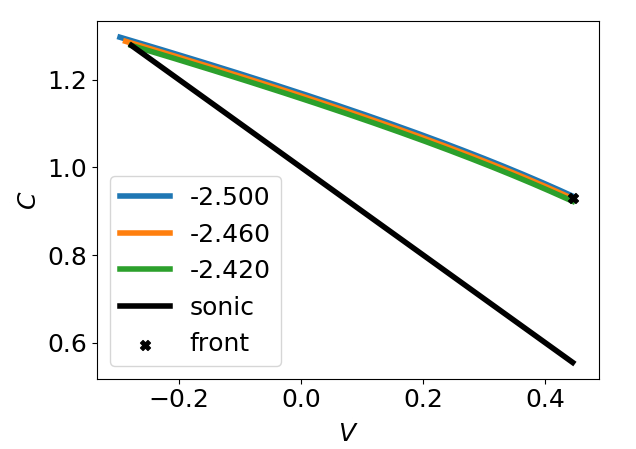}
\end{subfigure}
\begin{subfigure}{.5\textwidth}
  \centering
  \includegraphics[width=1\linewidth]{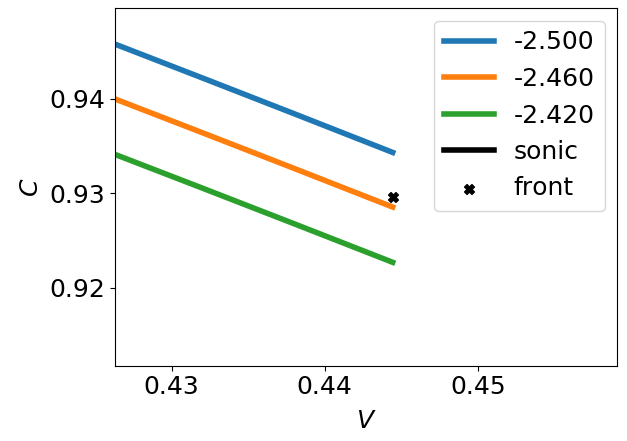}
\end{subfigure}

\caption{First type of integration, for $C_s > 0$. Here $\gamma = 3.5$ and $\omega = 3$. The numbers in the legend show the values of $\delta$ used in each integration. This integration is straightforward and there are no coordinate singularities.}
\label{fig:case1}
\end{figure}

The correct value of $\delta$ is such that the curve $C\left(V\right)$ satisfies both boundary conditions at the sonic point and the shock front. We use the shooting method to find the value of $\delta$. We guess a value for $\delta$, numerically integrate equation \ref{eq:dCdV_raw} w.r.t to $V$ from the sonic point to the shock front, and note the value of $C$ at the shock front. We then use the bisection method to refine the value of $\delta$ to minimise the distance between the value of $C$ obtained from numeric integration and its theoretical value (equation \ref{eq:C_final}). An example for some hydrodynamic $C\left(V\right)$ trajectories for the same value of $\gamma, \omega$ but different values of $\delta$ can be seen in figure \ref{fig:case1}.

\begin{figure}
\centering
\begin{subfigure}{.45\textwidth}
  \centering
  \includegraphics[width=1\linewidth]{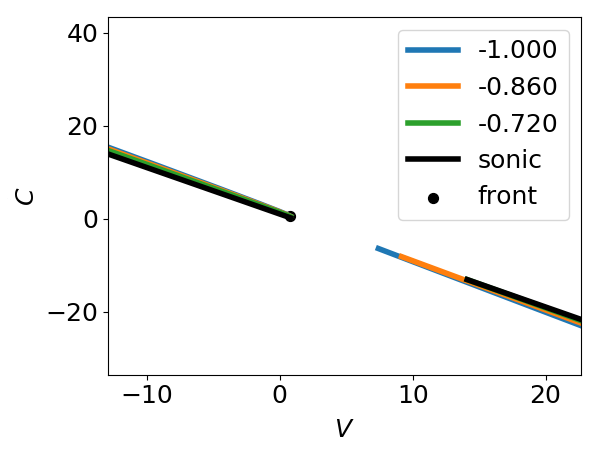}
\end{subfigure}
\begin{subfigure}{.45\textwidth}
  \centering
  \includegraphics[width=1\linewidth]{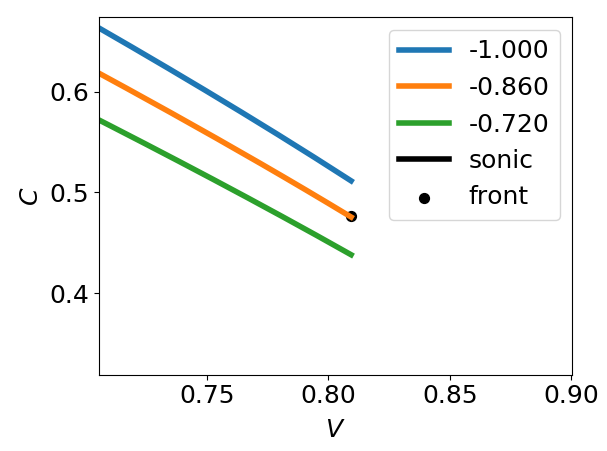}
\end{subfigure}

\caption{Second type of integration, for $C_s < 0$, and $\gamma = 1.47, \omega = 0.3$. In this situation, the shock and sonic points are on opposite sides of the boundary separating the fluid and the vacuum, so a mathematical singularity occurs when integrating from the sonic point to the shock front.}
\label{fig:case2}
\end{figure}

In the case $C_s>0$, numerical integration is straightforward since both the sonic point and the shock front are on the same side of the impact site. An example for several hydrodynamic trajectories for fixed $\gamma, \omega$ such that $C_s > 0$ can be seen in figure \ref{fig:case1}. The case $C_s < 0$ is more complicated because the integration goes through $x=0$. Because of the way we defined the self similar variables (equation \ref{eq:self_similar_vars}), they diverge at $x=0$, even though the physical quantities $v$ and $c$ remain finite. We can circumvent this difficulty by noting that the Mach number, i.e. the ratio between the velocity and the speed of sound, remains finite and changes continuously across $x=0$. The sonic point in the case  $C_s < 0$ lies in the fourth quadrant of the of the $C-V$ plane (i.e. positive $V$ and negative $C$). Numerical integration proceeds to even higher values of $V$, moving away from the shock front. Far away from the sonic point, the curve attains some asymptotic slope, and this slope is the Mach number at $x=0$, which we'll denote by $\mathcal{M}_0$. Once we have this piece of information, we can restart the numerical integration in the second quadrant ($C$ is positive and $V$ is negative), beginning from some arbitrarily highly negative $V=V_r$ (subscript r for restart) and $C_r = V_r/\mathcal{M}_0$. From this point we can continue the numerical integration all the way to the shock front. An example for several hydrodynamic trajectories for fixed $\gamma, \omega$ such that $C_s < 0$ can be seen in figure \ref{fig:case2}.

\begin{figure}
\centering
\begin{subfigure}{.5\textwidth}
  \centering
  \includegraphics[width=1\linewidth]{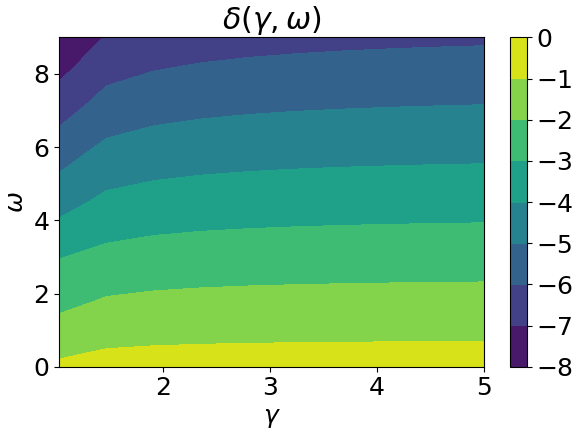}
\end{subfigure}
\begin{subfigure}{.5\textwidth}
  \centering
  \includegraphics[width=1\linewidth]{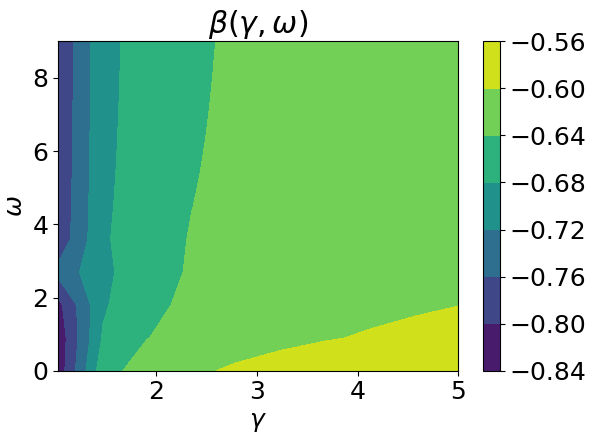}
\end{subfigure}
\caption{Contour plots of $\delta(\gamma, \omega)$ and $\beta(\gamma, \omega)$.
}
\label{fig:delta_beta}
\end{figure}

In this manner we calculated $\delta$ for different values of $\gamma$ and $\omega$. The results are displayed in the top panel of figure \ref{fig:delta_beta}. This figure shows that $\delta$ is mostly dependent on $\omega$, and weakly dependent on $\gamma$. For completeness, the asymptotic cases of $\gamma \rightarrow \infty$ ; $\omega \rightarrow \infty$ ; and  $\omega \rightarrow \infty,\gamma \rightarrow \infty$ are covered in the appendix.

This result prompted us to consider another power law index, $\beta = d \ln v / d \ln m$, which relates the velocity of the shock to the swept up mass. This parameter can be related to $\delta$ via
\begin{equation}
    \beta = \frac{\delta}{1+\omega} \label{eq:delta_beta_relation}
\end{equation} 
A contour plot of $\beta(\gamma, \omega)$ is shown in figure \ref{fig:delta_beta}, in which one can see that $\beta$ does not vary much as a function of $\gamma$ and $\omega$.

Conservation of energy and momentum laws allows us to set upper and lower limits on the power law parameters \citep{Zeldovich1967PhysicsPhenomena}
\begin{equation}
-\frac{1}{2}>\beta>-1  \label{eq:beta_limits}
\end{equation}
\begin{equation}
    -\frac{\omega + 1}{2} > \delta > -\omega-1 \label{eq:delta_limits}
\end{equation}
and \begin{equation}
    \frac{2}{3+\omega} > \alpha > \frac{1}{2+\omega} \, .
\end{equation}

\subsection{Comparison with Simulations}

In order to verify our results, we compare them to the numerical simulations performed in \citep{Yalinewich2019OpticalSurface}. The authors used hydrodynamic simulations to obtain the original power law parameter $\alpha$, defined in \ref{eq:shock_power_law}. They obtained $\alpha \left(\gamma = 5/3, \omega=3\right) = 0.19$ and $\alpha\left(\gamma=5/3, \omega=3/2\right) = 0.25$. We note that these calculations were performed in spherical geometry, so in order to calculate $\beta$ we use
\begin{equation}
    \beta = \frac{\delta}{3+\omega} = \frac{1-1/\alpha}{3+\omega} \, . \label{eq:beta_spherical}
\end{equation}
Using equation \ref{eq:beta_spherical} we calculate the numerical values $\beta \left(\gamma=5/3, \omega=3\right) = -0.71$ and $\beta \left(\gamma=5/3, \omega=3/2\right) = -0.67$. Using the numerical integration methods, we obtained $\beta \left(\gamma=5/3, \omega=3\right) = -0.671$ and $\beta \left(\gamma=5/3, \omega=3/2\right) = -0.663$. Despite using different geometric setups and approaches to determining the power law scaling parameters, the values for $\beta$ obtained here are reasonably close to those obtained previously with hydrodynamic simulations.

\section{Applications} \label{sec:app}

A number of previous studies have already considered the role of shock waves from a surface explosions in different astrophysical phenomena. \citep{Yalinewich2019AtmosphericImpacts} studied atmospheric mass loss from planetary collisions. \citep{Yalinewich2019OpticalSurface} studied the optical transient that results from a rapid release of energy by a neutron star while inside the envelope of a giant companion star. Finally, \citep{Yalinewich2021CraterImpacts} showed that the shapes of craters due to collisions with primordial black holes would differ from typical craters, and hence the moon can serve as a dark matter detector.

In addition to these cases, in this section we present more possible astrophysical applications to the formalism developed here. In particular, we discuss how the analytical formalism presented above can reproduce results from numerical hydrodynamic simulations and observations.

\subsection{Bolides}

Even when objects burn up completely in the atmosphere, they produce a pressure wave that can cause damage on the ground. The most recent example is the Chelyabinsk event \citep{Emelyanenko2013Astronomical2013}. The damage caused by the explosion (measured by the reports on damage per capita) declined with distance according to $r^{-2.4}$, similar to the measured decline in over-pressure $r^{-2.6}$. We can interpret this decline in pressure using the model developed above. After the shock wave reaches the ground, it expands further only in the horizontal direction. For this reason the swept up mass scales with the area $m \propto r^2$. The density at ground level is roughly constant, and so the pressure scales as $p \propto \rho v^2 \propto m^{-2 \beta} \propto r^{-4 \beta}$. Choosing $\beta = 2/3$, we find $p \propto r^{-2.7}$, which is relatively close to the empirical result. 

The damage caused by the shock depends on whether the shock is strong by the time it reaches the ground, i.e. whether the shock pressure is much larger than the atmospheric pressure when the shock reaches the ground.

As the bolide moves downward, it interacts with the atmosphere, decelerates and heats up. This heat causes it to expand, which increases the deceleration and heating rate and so forth. The air density at the the altitude at which the bolide decelerates considerably is \citep{Chyba1993TheAsteroid}
\begin{equation}
    \rho_d \approx \rho_i \left(\frac{R_i}{h}\right)^2
\end{equation}
where $\rho_i$ is the density of the impactor, $R_i$ its radius and $h$ is the scale height.
The bolide then deposits the energy in the gas and creates a shock wave that expands outward. At early times all parts of the shock interact with roughly the same density, so the shock evolves as a classical Sedov Taylor solution \citep{Taylor1950TheDiscussion, Sedov1946PropagationWaves}. The difference between density at different points on the shock becomes significant when the shock radius is comparable to the scale height. At that point the swept up mass is
\begin{equation}
    M_{ST} \approx \rho_d h^3 \approx \rho_i R_i^2 h \, .
\end{equation}
Since energy is conserved, the velocity at the end of the Sedov Taylor phase is
\begin{equation}
    v_s \approx v_i \sqrt{\frac{m_i}{M_{ST}}} \approx v_i \sqrt{\frac{R_i}{h}}
\end{equation}
where $m_i \approx \rho_i R_i^3$ is the mass of the impactor. As the shock moves to lower altitudes, it follows the solution found in the previous section, so the shock velocity on the ground is
\begin{equation}
    v_g \approx v_{ST} \left(\frac{M_{ST}}{\rho_0 h^3}\right)^{\beta} \approx v_i \sqrt{\frac{R_i}{h}} \left(\frac{R_i^2 \rho_i}{h^2 \rho_0}\right)^{\beta} \, .
\end{equation}
where $\rho_0$ is the air density on the ground so $\rho_0 h^3$ is the air mass enclosed within a radius $h$ from any point on the ground. The pressure on the ground is 
\begin{equation}
    p_g \approx \rho_0 v_g^2 \approx \rho_0 v_i^2 \frac{R_i}{h} \left(\frac{R_i^2 \rho_i}{ h^2 \rho_0}\right)^{2 \beta} \, .
\end{equation}
By comparing this pressure to the atmospheric pressure at ground level we can obtain a critical value for the impactor radius
\begin{equation}
    R_i \approx h \left(\frac{\rho_0}{\rho_i}\right)^{\frac{2 \beta}{4 \beta+1}} \left(\frac{p_0}{\rho_0 v_i^2}\right)^{\frac{1}{4 \beta+1}} \, .
\end{equation}
Substituting typical values and using $\beta = 2/3$ yields
\begin{equation}
    R_i \approx 66 \tilde{h} \tilde{\rho}_0^{1/11} \tilde{\rho}_i^{-4/11} \tilde{p}_0^{3/11} \tilde{v}_i^{-6/11} \, \rm m
\end{equation}
where $\tilde{h} = h / 8 \rm \, km$, $\tilde{\rho}_0 = \rho_0 / 10^{-3} \rm \, \frac{g}{cm^3}$, $\tilde{\rho}_i = \rho_i/ 3 \rm \, \frac{g}{cm^3}$, $\tilde{p}_0 = p_0 / 10^5 \, \rm Pa$ and $\tilde{v}_i = v_i/10 \, \rm \frac{km}{s}$. A impactor with a larger radius creates a shock that remains strong when it reaches the ground. In numerical simulations, the critical impactor size is around 50 metres \citep{Shuvalov2013AnAsteroids}. Incidentally, the over-pressure threshold for substantial damage to brick and concrete surfaces is also comparable to the ambient atmospheric pressure \citep{Glasstone1977TheEdition}. Therefore, if the shock wave is no longer strong by the time it reaches the ground, it is unlikely to cause substantial damage to buildings. Weak shocks can still cause damage to more brittle materials such as glass and wood, as was the case with the Chelyabinsk bolide.

Finally, it is possible to estimate the atmospheric mass loss from bolides. An impactor moving faster than the escape velocity will create a decelerating shock wave, so that after it has swept up a certain amount of mass the velocity will drop below the escape velocity. Since the velocity does not change considerably after the shock, the amount of mass lost in this way will be comparable to the mass accelerated to velocities exceeding the escape velocity $v_e$. Hence, we can approximate the amount of atmospheric mass loss $M$ by
\begin{equation}
    M \approx m_i \left(\frac{h}{R_i}\right)^{\frac{2 \beta-1}{\beta}} \left(\frac{v_i}{v_e}\right)^{1/\beta}\, .
\end{equation}
Verification of this relation and discussion of its implication for planet formation is relegated to a future work.

\subsection{Shoemaker Levy 9}

The impact of the fragments of the comet Shoemaker Levy 9 produced bright visible and infrared flashes that lasted, depending on the band, between tens of seconds and tens of minutes. A direct comparison between the observational data and theory is problematic because of a number of reasons. First, some of the fragments landed on the far side of Jupiter, so that only the spacecraft Galileo was able to observe the early phases of the impact \citep{Neukum1995ThePhase.}, while ground based instruments could only begin to observe after the plumes rose above the horizon (although some did detect a faint precursor due to the passage of the fragments through the tenuous upper layers of Jupiter's atmosphere). Second, some of the ground based telescopes became saturated and thus missed the peak \citep{Graham1995TheSL9}. Nevertheless, such a comparison has been performed \citep{Zahnle1995AImpact}, using a similar model for the shock propagation, but without the analytic justification. We will not repeat that analysis here, but instead focus on the analytic predictions that can be obtained from our formalism. Also, for simplicity in this section we assume that $\beta = 2/3$.

Most of the light observed after the impact of Shoemaker Levy 9 on Jupiter did not originate from the bolide itself, but from ejecta from the impact site that collided with more distant parts on the surface of Jupiter \citep{Zahnle1995AImpact}. The range of every fluid shell depends on the velocity with which it emerges from the basin
\begin{equation}
    r \approx \frac{v^2}{g} \approx \frac{R_{i}^{\frac{10}{3}} h^{\frac{2}{3}} \rho_{i}^{\frac{4}{3}} v_{i}^{2}}{g m^{\frac{4}{3}}} \, .
\end{equation}
where $g$ is the Jovian surface gravity. The time it takes a fluid shell to fall down to the surface is 
\begin{equation}
    t_f \approx \frac{v}{g} \approx \frac{R_{i}^{\frac{5}{3}} \sqrt[3]{h} \rho_{i}^{\frac{2}{3}} v_{i}}{g m^{\frac{2}{3}}} \, .
\end{equation}
At any moment an observer can only see fluid elements that had time to fall $t>t_f$. When the ejecta collides with the surface, it is shocked, and its temperature is given by
\begin{equation}
    T \approx \mu v^2 / k \approx \frac{g^{2} \mu t^{2}}{k} \, .
\end{equation}
where $\mu$ is the atomic mass and $k$ is the Boltzmann constant. The luminosity is given by blackbody emission
\begin{equation}
    L \approx \sigma T^4 r^2 \approx \frac{g^{10} \mu^{4} \sigma t^{12}}{k^{4}} \, .
\end{equation}
We find that the luminosity rises very steeply, and so attains its peak value almost instantaneously, in accordance with observations \citep{Chapman1996GalileoImpacts}. This peak luminosity comes from the fastest shell, which is the Sedov mass. The corresponding peak bolometeric luminosity is
\begin{equation}
    L_p \approx \frac{R_{i}^{12} \mu^{4} \sigma v_{i}^{12}}{g^{2} h^{12} k^{4}} \approx 10^{22} \tilde{R}_i^{12} \tilde{\mu}^4 \tilde{v}_i^{12} \tilde{g}^{-2} \tilde{h}^{-12} \, \rm erg/s
\end{equation}
where $\tilde{R}_i = R_i/24 \, \rm km$, $\tilde{v}_i = v_i / 60 \rm \, km/s$, $\tilde{g} = g/60 \rm m/s^2$, $\tilde{h} = h / 24 \, \rm km$, $\tilde{\mu} = \mu / m_p$ and $m_p$ is the proton mass. This result is a factor of a few larger than the observed luminosity \citep{Chapman1996GalileoImpacts}, however the results are very sensitive to the parameters. The duration of the transient will be comparable to the fallback time
\begin{equation}
    t_p \approx \frac{v_i R_i}{g h} \approx 40 \tilde{v}_i \tilde{R}_i \tilde{g}^{-1} \tilde{h}^{-1} \, \rm s.
\end{equation}
This estimate is also in agreement with the observations.
The temperature at the peak is given by
\begin{equation}
    T_p \approx \frac{R_{i}^{2} \mu v_{i}^{2}}{h^{2} k} \approx 750 \tilde{R}_i^2 \tilde{\mu} \tilde{v}_i^2 \tilde{h}^{-2} \, \rm K \, .
\end{equation}
The observed temperature is larger than our result by a factor of three  \citep{Nicholson1995PalomarCurves}.

\subsection{Asteroid Impact Avoidance}

The impact of rocky bodies on Earth poses a danger to the life and safety of its inhabitants. For this reason, over the years different strategies for asteroid impact avoidance have been studied \citep[and references therein]{NationalResearchCouncil2010DefendingReport}. Asteroids whose diameter is less than a few hundred meters can be deflected using kinetic impactors (sometimes called mass drivers). The net change in the momentum of the target is larger than just the momentum of the impactor, since upon impact debris from the target fly in the opposite direction. We can use the formalism above to estimate the net change in the momentum of the target.

Let us consider an impactor of mass $m$ that collides with a much larger target at a velocity $v_i$. As a result, a shock wave emerges from the impact site and excavates a crater. Crater excavation stops when the shock velocity drops below some critical velocity $v_e$, which is comparable to the velocity with which elastic waves propagate. The mass of the excavated material is
\begin{equation}
    M \approx m \left(\frac{v_i}{v_e}\right)^{1/\beta} \, .
\end{equation}
Most of the ejected material moves with a velocity $v_e$, so the net change in momentum is $M v_e$. For convenience, we normalise this change in target momentum with the momentum of the incident impactor
\begin{equation}
    \xi \approx \frac{M v_e}{m v_i} \approx \left(v_i/v_e\right)^{\frac{1-\beta}{\beta}} \, .
\end{equation}
For $\beta = 2/3$, we get $\xi \propto v_i^{0.5}$, which is similar to the behaviour found in numerical simulations $\xi \propto v_i^{0.44} $ \citep{BruckSyal2016DeflectionProperties}.

Larger asteroid can only be deflected using nuclear surface explosions \citep{Ahrens1992DeflectionAsteroids}. We can use our formalism to relate the energy and depth of the explosion to the net momentum change. If the initial depth is $h$ and the ambient density of the asteroid is $\rho_0$, then the shock wave conserves energy until it sweeps up a mass $\rho_0 h^3$, at which point the shock velocity is $\sqrt{E/\rho_0 h^3}$. At later times the shock evolves according to our new formalism. Thus, the mass of the ejecta is
\begin{equation}
    M \approx \rho_0 h^3 \left(\sqrt{\frac{E}{\rho_0 h^3 v_e^2}}\right)^{1/\beta}
\end{equation}
where $v_e$ is the escape velocity from the asteroid. The net change momentum is
\begin{equation}
    I \approx v_e M \approx v_e \rho_0 h^3 \left(\sqrt{\frac{E}{\rho_0 h^3 v_e^2}}\right)^{1/\beta} \propto h^{3-3/2\beta} \, .
\end{equation}
If $\beta =2/3$, we have $I \propto h^{0.75}$, which is similar to the predictions based on empirical scaling laws \citep{Meshcheryakov2015EstimatedImpact}. If the explosion is too deep, then by the time the explosion makes it to the surface the shock velocity drops below the escape velocity and the net change in momentum is zero. The optimal depth is determined by the condition that the shock is at the escape velocity when it reaches the surface
\begin{equation}
    h_{o} \approx \left(\frac{E}{\rho_0 v_e^4}\right)^{1/3} \, .
\end{equation}

\subsection{Hypervelocity White Dwarfs}
Recent observations revealed an intriguing class of celestial objects. These are white dwarfs with velocities that are markedly different than the other stars in their neighbourhoods, and also feature unusual surface composition \citep{Kepler2016AAtmosphere, Shen2018Supernovae}. It has been suggested that these are white dwarfs that underwent some episode of runaway thermonuclear burning that stopped before it consumed the entire white dwarf \citep{Gansicke2020SDSSIgnition}. If the reaction did not stop, then the white dwarf would have explode as a type Ia supernova and leave no remnant behind. For this reason, a partially burned white dwarf is sometimes referred to as a supernova survivor. It has also been suggested that these events give rise to subluminous supernovae \citep{Foley2013TypeExplosion}.

Many models have been proposed to explain type Ia supernova \citep{Hillebrandt2000TypeModels}, and in most of them, the explosion is initiated at a point that is offset from the centre. Numerical simulations have shown that sometimes the reaction will only run through a portion of the white dwarf rather and not destroy it entirely \citep{Jordan2012Failed-DetonationCores}. We can use the formalism developed here to calculate the aftermath of such a failed supernova. For simplicity, let us consider the white dwarf to be a sphere of uniform density $\rho$. Suppose further that the runaway reaction has burned through a hemisphere of radius $h$ below the surface. The mass of the burnt material is therefore $\rho h^3$. We also assume that energy released per unit mass is $\varepsilon$, so the burnt material is accelerated to a velocity $\sqrt{\varepsilon}$. From that hot spot a shock wave emerges and travels through the white dwarf, but does not ignite more material. If the escape velocity from the white dwarf is $v_e$, then the net mass loss is 
\begin{equation}
    \Delta M \approx \rho h^3 \left(\frac{\varepsilon}{v_e^2}\right)^{1/2\beta} \, .
\end{equation}
The mass ejected from the explosion moves at the escape velocity, so the velocity kick to the survivor is
\begin{equation}
    \frac{\Delta v}{v_e} \approx \left(\frac{h}{R}\right)^3 \left(\frac{\varepsilon}{v_e^2}\right)^{1/2\beta}
\end{equation}
where $R$ is the radius of the white dwarf. For $\beta = 2/3$ we find
\begin{strip}
\begin{equation}
    \frac{\Delta M}{M_{WD}} \approx \left(\frac{h}{R}\right)^3 \left(\frac{\varepsilon}{v_e^2}\right)^{3/4} \approx 0.05 \left(\frac{h/R}{0.3}\right)^3 \left(\varepsilon /\frac{{\rm MeV}}{m_p}\right)^{0.75} \left(\frac{v_e}{6 \cdot 10^3 \, \rm \frac{km}{s}}\right)^{-1.5}
\end{equation}
\end{strip}
where $M_{WD}$ is the mass of the white dwarf and $m_p$ is the proton mass. The kick velocity is
\begin{equation}
    \Delta v \approx 340 \left(\frac{h/R}{0.3}\right)^3 \left(\varepsilon / \frac{\rm MeV}{m_p}\right)^{0.75} \left(\frac{v_e}{6 \cdot 10^3 \, \rm \frac{km}{s}}\right)^{-0.5} \, \rm \frac{km}{s} \, .
\end{equation}
These values are similar to those found in numerical simulations \citep{Jordan2012Failed-DetonationCores}.

\section{Conclusion} \label{sec:conclusion}

In this paper, we develop a self similar model for shocks propagating down stellar or planetary atmospheres with graded density profiles as a result of a rapid, localised deposition of energy. We achieve this by generalising the impulsive piston problem to account for a graded density profile. In this model, instead of considering the full three dimensional problem, we consider a one dimensional slab symmetric analogue. In the analogue problem, a thin wafer hits a half space target whose density is allowed to vary in the direction normal to the surface. The evolution of the resulting shock is determined by the conditions at the sonic point, a singularity beyond which information cannot travel to the shock front. 

Under these assumptions, we can apply the self similar method to solve the hydrodynamic equations \citep{Landau1987FluidMechanics}. We do so by reducing the partial differential flow equations to a single ordinary differential equation. We determine the deceleration parameter of the shock $\beta = d \ln \dot{X} / d \ln m$ (where $\dot{X}$ is the shock velocity and $m$ is the swept up mass) by imposing boundary conditions at the shock front and at the sonic point, and using the shooting method.

We find that the parameter $\beta$ is weakly dependent on the density profile. For this reason, we can use this scaling relation even for different geometries, since the main difference between slab, cylindrical and spherical shock is the relation between the swept up mass and the distance from the centre of the explosion. 

We also demonstrate that the method we developed can be used to study a wide range of astrophysical scenarios. These scenarios include the transient resulting from the impact of meteors on Earth and other planets, mass loss due to giant planetary collisions, crater morphology, explosions in a common envelope and failed type Ia supernovae.

\section*{Acknowledgements}

We would like to thank Jonathan Fortney for suggesting the title for the paper. AY is supported by the Natural Sciences and Engineering Research Council of Canada (NSERC), funding
reference \#CITA 490888-16. This work made use of the sympy \citep{Meurer2017SymPy:Python}, numpy \citep{Oliphant2015GuideNumPy} and matplotlib \citep{Hunter2007Matplotlib:Environment} python packages.

\section*{Data Availability}

The equations and results for Section 2 and the appendix can be found in: \href{https://github.com/aremorov/stellar_atm_shocks}.

\bibliographystyle{mnras}
\bibliography{main}

\appendix
\section{Asymptotic Behaviour}

\subsection{Asymptotic Behaviour for an Infinite Adiabatic Index}
In the limit $\gamma \rightarrow \infty$, the equations used in Section 2.2 break down. This is because the integration interval $V \in \left[\frac{2\delta + \omega}{\delta (2-\gamma)+\omega}, \frac{2}{\gamma+1}\right]$ shrinks to zero width in this limit. To overcome this difficulty, we define a new variable $W = \gamma V$, as was done in  \citep{Yalinewich2020SelfCollisions}. In this new variable, the integration domain becomes $W \in \left[-2-\frac{\omega}{\delta},2\right]$, and so remains finite in the limit $\gamma\rightarrow \infty$. The boundary conditions for $C$ simplifies as well, and become: $C = 1$ at the sonic point and $C = \sqrt{2}$ at the shock front. The expression for the derivative involving the new variable is given by

\begin{equation}
    \frac{d C}{d W} = \frac{C (W \delta + C^{2} \omega + 2 C^{2} + 2 \delta - 2)}{2 (W C^{2} + W \delta - W + 2 C^{2} \delta + C^{2} \omega)} \, 
\end{equation}

At the sonic point, the slope is given by

\begin{equation}
    \left. \frac{d C}{d W} \right|_s = - \frac{\delta (\delta - \omega + \sqrt{9 \delta^{2} + 2 \delta \omega - 12 \delta + \omega^{2} + 4} - 2)}{8 \delta^{2} + 4 \delta \omega - 8 \delta - 4 \omega}\, 
\end{equation}

\begin{figure}
    \centering
    \includegraphics[width=0.9\linewidth]{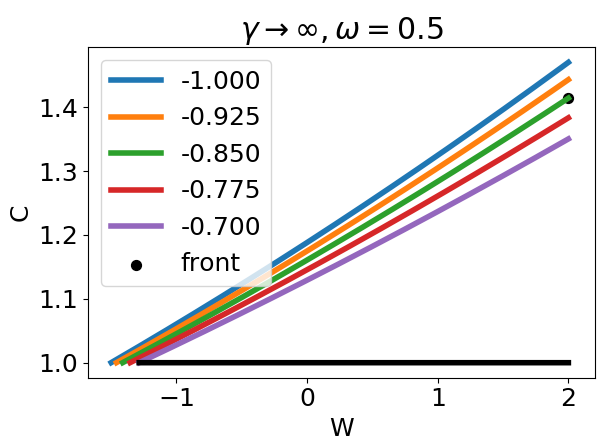}
\caption{Hydrodynamic trajectories for the case $\gamma \rightarrow \infty, \omega = 0.5$. Different coloured lined represent trajectories with different values of $\delta$. The sonic line is represented by the black line, and the shock front by the black cross.}
    \label{fig:gamma_infinity_trajectory}
\end{figure}

\begin{figure}
\centering
\begin{subfigure}{.45\textwidth}
  \centering
  \includegraphics[width=1\linewidth]{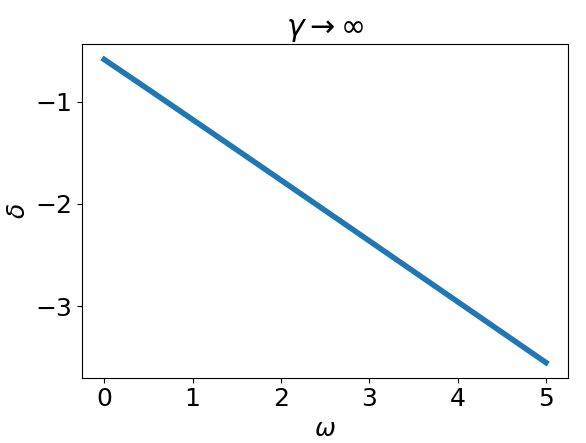}
\end{subfigure}
\begin{subfigure}{.45\textwidth}
  \centering
  \includegraphics[width=1\linewidth]{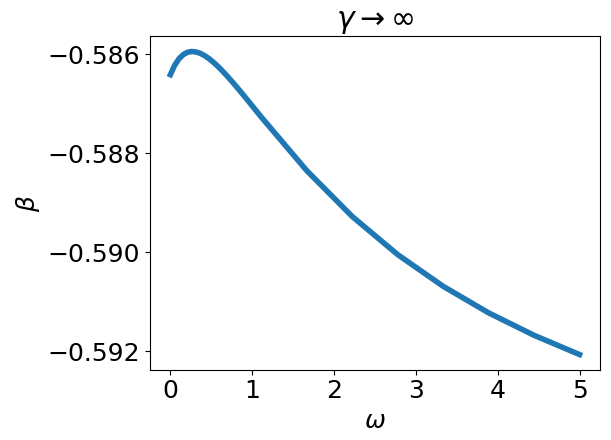}
\end{subfigure}

\caption{Using the shooting integration method with a variety of $\omega$, plots of $\delta(\omega)$ (at the top) and $\beta(\omega)$ (at the bottom) for $\gamma\rightarrow\infty$ can be obtained. Although $\delta$ appears to be linearly dependent on $\omega$, this is actually not the case, since $\frac{d\delta}{d\omega}$ is not a constant function of $\omega$. It is noted that $\beta$ has a much lower dependence on $\omega$ than $\delta$.}
    \label{fig:gamma_infinity_several}
\end{figure}

We can determine $\delta(\omega)$ and $\beta(\omega)$ using the shooting integration method described in the previous section, since we know the boundary conditions and $\frac{dC}{dW}$. A set of sample trajectories, for $\omega = 0.5$, is shown in figure \ref{fig:gamma_infinity_trajectory}. Applying the shooting integration method with varying $\omega$ allows one to obtain the dependencies of $\delta$ and $\beta$ on $\omega$, shown in figure \ref{fig:gamma_infinity_several}. The values of $\beta$ and $\delta$ in this asymptotic case fall below their respective energy conservation limits in equations \ref{eq:beta_limits}, \ref{eq:delta_limits}, unlike in the opposite limit $\gamma = 1$, where the values of these parameters are equal to the momentum conservation limits.

\subsection{Asymptotic behaviour for an Infinite Density Slope}
The limit $\omega \rightarrow \infty$ is equivalent to an exponential density profile. In this scenario, different equations for the  density profile, dimensionless hydrodynamic quantities and shock front trajecory have to be used. A similar problem has been considered \citep{Schlichting2015AtmosphericImpacts}, so many of the expressions are similar. We note however that \citep{Schlichting2015AtmosphericImpacts} considered the portion of the shock wave that travels in the opposite direction, towards lower densities, and so their power law index is different.

The density profile is defined by $\rho(x) = \rho_0 \exp\left(\frac{x}{h}\right)$ where $\rho_0$ is the density at a reference point (where the explosion started), x is the distance from the reference point, and h is the atmospheric scale height. In our situation, the shock wave gets slower as time progresses, while in \citep{Schlichting2015AtmosphericImpacts} it gets faster and faster and traverses an infinite distance after a finite time.
The dimensionless position is $\chi = \frac{X(t)-x}{h}$ and the dimensionless hydrodynamic variables are
\begin{equation}
    v(x,t) = \frac{d X(t)}{dt}  V  \quad \quad
    c(x,t) = \frac{d X(t)}{dt}  C \quad \quad
    \rho(x,t) = \rho_0 e^{X(t)/h} D \label{eq:self_similar_vars2}
\end{equation}

In the new framework, we start again with the same slab symmetric hydrodynamic equations \ref{eq:mass_conservation}, \ref{eq:momentum_conservation} and \ref{eq:entropy_conservation} and ideal gas equation of state, and then substitute in the new dimensionless parameters defined in \ref{eq:self_similar_vars2}.

Like in section 2.2, after these substitutions, the equations contain terms involving different time derivatives of $X$. The time derivatives can be factored out using equation, by eliminating the second derivative using
\begin{equation}
    \frac{d^2 X}{d t^2} = -\frac{\delta}{h} \left(\frac{d X}{d t}\right)^2 \,.
\end{equation}

This equation looks different from equation \ref{eq:X_derivatives} because the trajectory of the shock front is not a power law of time, but instead has a logarithmic dependence on time given by $X(t) = \frac{h}{\delta} \ln(1 + t/t_0$) where h is the atmosphere scale height and $t_0$ is a scale time (equation 11 in \citep{Schlichting2015AtmosphericImpacts}).

In this limit, the hydrodynamic equations equations \ref{eq:mass_conservation}, \ref{eq:momentum_conservation} and \ref{eq:entropy_conservation}
can be represented as a matrix equation $\overleftrightarrow{M} d\vec{A}/\chi = \vec{B}$ where:

\begin{equation}
M = \begin{bmatrix}1 - V & - D & 0\\- C^{2} & \gamma (1 - V) D & - 2 C D\\- \gamma C - (- \gamma C + C) V + C & 0 & - 2 D V + 2 D\end{bmatrix}
\end{equation}

\begin{equation}
\vec{A} = \begin{bmatrix} D\\ V\\ C\end{bmatrix}
\end{equation}

\begin{equation}
\vec{B} = \begin{bmatrix}- D\\- \gamma \delta D V\\\gamma C D - C D - 2 \delta C D\end{bmatrix}
\end{equation}

Once again, these equations are linear in the derivatives $\frac{dV}{d\chi}, \frac{dC}{d\chi}, \frac{dD}{d\chi}$, so we can isolate them, and obtain a single differential equation of $\frac{dC}{dV}$ by going through the same process as was done for equations 11 to 14:

\begin{equation}
\frac{dC}{dV} = \frac{C ( \frac{C^2 \gamma}{\delta}  - \frac{C^2}{\delta} - 2 C^2 - V^2 \gamma^2 + 3 V^2 \gamma + V \gamma^2 - 5 V \gamma + 2 \gamma)}{- \frac{2 C^2 V}{\delta} - 4 C^2 V + \frac{2 C^2}{\delta}+ 4 C^2 + 2 V^3 \gamma - 4 V^2 \gamma + 2 V \gamma} \label{eq:dCdV_raw2}
\end{equation}

Like in section 2.3, in order integrate equation \ref{eq:dCdV_raw2}, boundary conditions and the value of $\alpha$ are needed. Once again, the boundary conditions at the shock front are:

\begin{equation}
    V_f = \frac{2}{\gamma+1} \label{eq:v_shock_front2}
\end{equation}
and
\begin{equation}
    C_f = \frac{\sqrt{2 \gamma \left(\gamma-1\right)}}{\gamma+1} \, .  \label{eq:C_final2}
\end{equation}

The denominator of equation \ref{eq:dCdV_raw2} vanishes on the sonic line, given by equation \ref{eq:sonic_line}. At the sonic point, the numerator also vanishes. This allows us to obtain the sonic point boundary conditions for this scenario:

\begin{equation}
V_s = \frac{\frac{1}{\delta}+2}{\frac{1}{\delta}+2-\gamma} \label{eq:v_sonic2}
\end{equation}
\begin{equation}
C_s = -\frac{\gamma}{\frac{1}{\delta}+2-\gamma}  \label{eq:c_sonic2}
\end{equation}

\begin{figure}
\centering
\begin{subfigure}{.45\textwidth}
  \centering
  \includegraphics[width=1\linewidth]{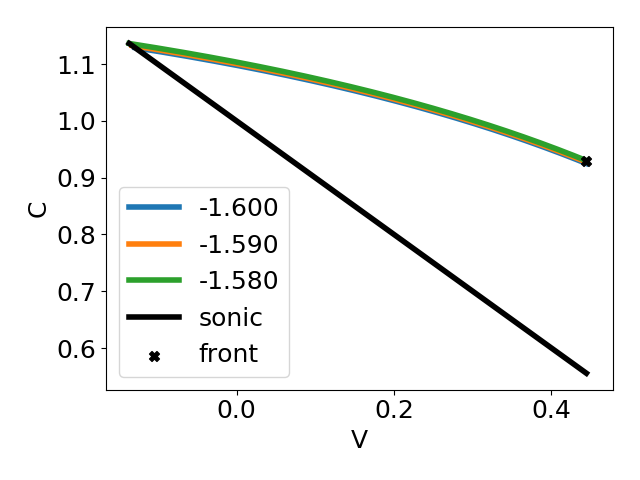}
\end{subfigure}
\begin{subfigure}{.45\textwidth}
  \centering
  \includegraphics[width=1\linewidth]{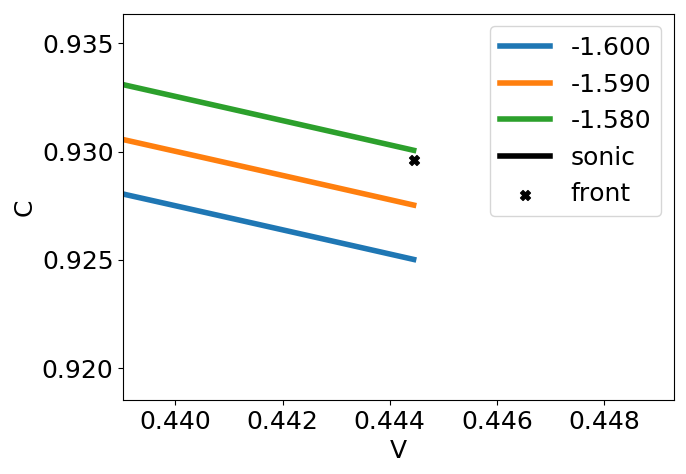}
\end{subfigure}

\caption{Hydrodynamic trajectories for the exponential density profile and $\gamma = 3.5$. Fortunately, in this scenario, the integration is always straightforward and there are no mathematical singularities. Once a $\frac{1}{\delta}$ value is found whose integration curve ends close enough to the shock front boundary condition, its reciprocal is taken to obtain the corresponding $\delta$ value.}
\label{fig:omega_infinity_trajectories}
\end{figure}

\begin{figure}
    \centering
    \includegraphics[width=0.9\linewidth]{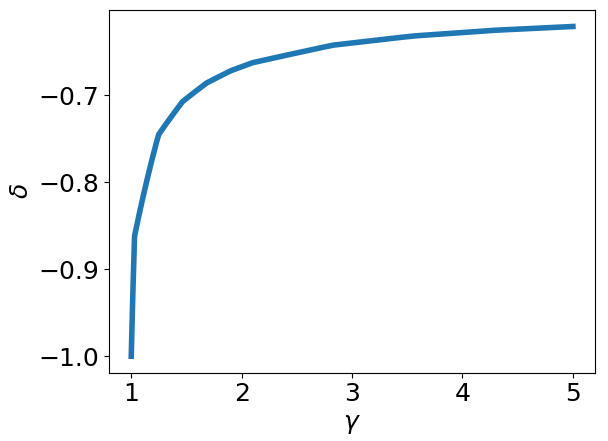}
    \caption{ A plot of $\delta(\gamma)$ for the exponential density profile scenario ($\omega \rightarrow \infty$) can be obtained by applying the shooting integration method with a variety of $\gamma$ and taking the reciprocal of the respective $\frac{1}{\delta}$ values.}
    \label{fig:d(y)_omega_inf}
\end{figure}

Knowing the boundary conditions and $\frac{d C}{d V}$, we can integrate from the sonic point to the shock front. The integration process, with the exception of the exact formulas for $\frac{dC}{dV}$ and boundary conditions, is equivalent to section 2.3 above, so the details will not be repeated here. However, one thing to note is that the $C_{s} < 0$ never arises, because $\frac{1}{\delta} + 2 < \gamma$ for all physically possible ranges of $\delta, \gamma$. This means that only the first integration case, where the sonic point and shock front are on the same side of the impact site, occurs. An example of a hydrodynamic trajectory for this scenario is shown in figure \ref{fig:omega_infinity_trajectories}. The bisection solving method can be used to obtain $\delta$ for each possible value of $\gamma$, and then taking the reciprocal gets $\delta$. This relation $\delta(\gamma)$ is plotted in figure \ref{fig:d(y)_omega_inf}.  In the limit of $\gamma = 1$, we get $\delta = -1$, which corresponds to momentum conservation. In the limit $\gamma \rightarrow \infty$, which corresponds to energy conservation, the solution converges to approximately $\delta=-0.597$. Obtaining this value requires the same substitution $W = \gamma V$ present in section A1.

\subsection{Asymptotic behaviour for Infinite Adiabatic Index and Density Slope}
Once again, when $\gamma \rightarrow \infty$, the integration along $V \in \left[\frac{\frac{1}{\delta}+2}{\frac{1}{\delta}+2-\gamma}, \frac{2}{\gamma+1}\right]$ shrinks to zero width, and using the new variable $W = \gamma V$ instead of $V$ remedies this. Now the integration range for $W$ is $\left[-2-\frac{1}{\delta},2\right]$, and this is not empty, so the integration process can occur. The boundary conditions in $C$ are $C = 1$ at the sonic point and $C = \sqrt{2}$ at the shock front. The expression for the derivative in this situation is:

\begin{equation}
    \frac{d C}{d W} = \frac{C (W +  \frac{C^{2}}{\delta} + 2)}{2 (W +  \frac{C^{2}}{\delta} + 2 C^{2})} \, 
\end{equation}

And at the sonic point this slope is:

\begin{equation}
    \left. \frac{d C}{d W} \right|_s = - \frac{\frac{3}{\delta} + \sqrt{\frac{9}{\delta^{2}} + \frac{8}{\delta} +16}}{4( \frac{1}{\delta} +2)}\, 
\end{equation}

\begin{figure}
    \centering
    \includegraphics[width=0.9\linewidth]{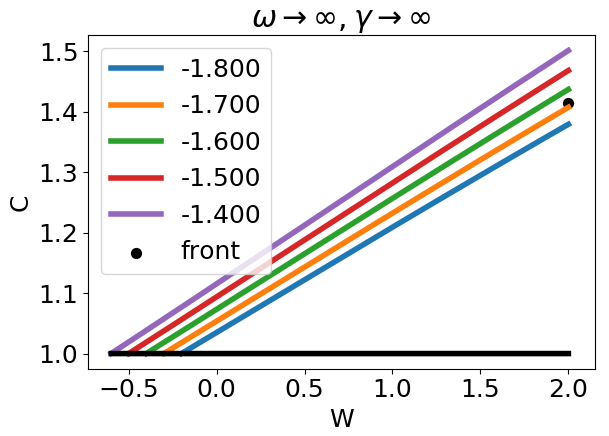}
\caption{Hydrodynamic trajectories for the case $\omega \rightarrow \infty, \gamma \rightarrow \infty$. Different coloured lined represent trajectories with different values of $\frac{1}{\delta}$. The sonic line is represented by the black line, and the shock front by the black cross.}
    \label{fig:both_limits}
\end{figure}

We can now determine $\delta(\gamma)$ using the shooting integration method, since the boundary conditions and $\frac{dC}{dW}$ are known. Some hydrodynamic trajectories for different values of $\delta$ close to the solution are shown in figure \ref{fig:both_limits}. Using a bisection solver, we find that $\lim_{\gamma\rightarrow \infty}\delta\approx-0.597$.

\bsp	
\label{lastpage}
\end{document}